# Reliable Annotations with Less Effort: Evaluating LLM-Human Collaboration in Search Clarifications

Leila Tavakoli  
Services Australia  
Melbourne, Australia  
leila.tavakoli31@gmail.com

Hamed Zamani  
University of Massachusetts Amherst  
Amherst, United States  
zamani@cs.umass.edu

## Abstract
Despite growing interest in using large language models (LLMs) to automate annotation, their effectiveness in complex, nuanced, and multi-dimensional labelling tasks remains relatively underexplored. This study focuses on annotation for the search clarification task, leveraging a high-quality, multi-dimensional dataset that includes five distinct fine-grained annotation subtasks. Although LLMs have shown impressive capabilities in general settings, our study reveals that even state-of-the-art models struggle to replicate human-level performance in subjective or fine-grained evaluation tasks. Through a systematic assessment, we demonstrate that LLM predictions are often inconsistent, poorly calibrated, and highly sensitive to prompt variations. To address these limitations, we propose a simple yet effective human-in-the-loop (HITL) workflow that uses confidence thresholds and inter-model disagreement to selectively involve human review. Our findings show that this lightweight intervention significantly improves annotation reliability while reducing human effort by up to 45%, offering a relatively scalable and cost-effective yet accurate path forward for deploying LLMs in real-world evaluation settings.

## 1 Introduction

Cranfield experiments established the foundation for offline evaluation in Information Retrieval (IR), setting standards for experimental design and performance evaluation [36]. Based on this paradigm, traditional IR evaluation is heavily based on human annotators for relevance labelling and quality assessment, and has shaped the field for decades, well before the rise of modern web search engines [77]. Although fundamental, these human-based evaluations are costly, slow, and difficult to scale. The recent emergence of large language models (LLMs) offers promising avenues for automating or partially automating these annotation processes. In fact, several studies demonstrate that LLMs can achieve near-human performance in certain annotation tasks [27, 28, 84], positioning them as potential replacements or enhancements to manual labelling efforts [8, 61, 73]. However, LLMs are not yet reliable enough for full autonomy, especially in nuanced or high-stakes tasks such as medical triage [32], legal interpretation [57], or fine-grained relevance judgements [50, 55]. Their behaviour can be brittle: sensitive to prompt phrasing, temperature settings, and domain shifts, and their confidence calibration is often unreliable. Consequently, recent studies advocate for Human-in-the-Loop (HITL) approaches, combining the efficiency and scalability of LLMs with human oversight (if and when needed) to ensure annotation quality and reliability [59].

In this study, we conduct a holistic evaluation and propose a practical yet effective HITL pipeline tailored for annotation tasks that involve multiple dimensions of complexity.[1] Rather than simply presenting another HITL framework, we explicitly assess the extent to which current state-of-the-art LLMs can realistically replace human annotators in *search clarification* as a complex task in (conversational) IR. We use the task of search clarification—which involves generating and evaluating clarification questions for ambiguous and faceted queries—as an example of complex evaluation task in (conversational) IR for which high-quality, multi-dimensional, and fine-grained human data is available, e.g., MIMICS-Duo [62]. In more detail, clarification tasks inherently require nuanced, subjective judgements across multiple dimensions (e.g., coverage, clarity, user preference), making them particularly suitable for evaluating the strengths and limitations of LLM-based annotation. Search clarification is broadly utilised by web search engines (e.g., Google's "refine your search") and conversational systems (e.g., Alexa, Siri), See Figure 1 for an example. However, current annotation and evaluation methods for clarification quality heavily rely on costly, large-scale human crowdworkers [62]. Our research explicitly examines how effectively a simple, modular HITL strategy—leveraging multiple LLMs, confidence-based filtering, and minimal human intervention—can reduce annotation effort and cost without compromising annotation quality.

To rigorously explore the capabilities of current LLMs, we evaluate four prominent proprietary as well as open-source LLMs: GPT-4o [2], Claude 3 [9], Cohere Command R [20], and Mistral 7B [34], under both zero-shot (ZSS) and few-shot (FSS) settings. Our focus on LLMs under ZSS, FSS, and HITL systems, instead of fully supervised models, reflects a deliberate effort to reduce the time and resources needed for human labelling. Using the same detailed guidelines provided to human annotators, we provide an empirical and transparent comparison between human annotators and LLM-generated annotations. Additionally, we systematically assess sensitivity to hyperparameters and prompts to offer deeper insights into LLM reliability. The novelty and significance of our research lie not merely in the technical components of the HITL framework but primarily in our comprehensive, critical assessment of whether and how LLMs can practically be deployed in subjective IR annotation workflows. Specifically, our main contributions are:

- Providing the first empirical evidence on automatic evaluation of search clarification tasks, highlighting current LLM limitations in nuanced, subjective annotation tasks.
- Demonstrating a clear and practical method—using aggregated LLM predictions, confidence thresholding, and minimal human intervention—to effectively identify and flag uncertain cases for human review and **reduce human annotation effort and cost by 24–45% with no considerable quality degradation**.
- Conducting a detailed prompt and hyperparameter sensitivity analysis, highlighting practical considerations for future research and deployment.

---

[1]We publicly release our data, code, prompts, and annotations at https://github.com/Leila-Ta/HITL-Framework-Search-Clarification.

## 2 Problem Statement

While LLMs have demonstrated considerable promise for automating annotation tasks [24, 54, 68], current research has predominantly focused on relatively straightforward, single-dimensional tasks. Consequently, a critical knowledge gap persists in understanding the capabilities and limitations of LLMs in handling complex, nuanced annotation tasks common across various IR scenarios. Such tasks typically require deep contextual understanding, subjective interpretation, and multi-dimensional assessment, making them inherently challenging for automated approaches.

One representative example of such nuanced tasks is search clarification [5, 6, 78], where systems generate questions to help users refine ambiguous or underspecified queries. Effective annotation of these clarifications demands balancing multiple subjective factors, including relevance, clarity, user engagement, and interpretability. However, despite being widely applicable to IR contexts, search clarification is just one instance within a broader class of annotation challenges that require nuanced judgements and sophisticated evaluation frameworks.

Our research directly addresses this broader gap by systematically evaluating the extent to which LLMs can replicate human-like annotations in complex, multi-dimensional annotation tasks, with search clarification serving merely as a representative testbed. Furthermore, we investigate how incorporating HITL principles can strategically mitigate the limitations of current LLMs, ensuring high-quality annotations while significantly reducing annotation costs and efforts. To explore this systematically, we define three distinct yet complementary annotation tasks that vary in complexity and cognitive load:

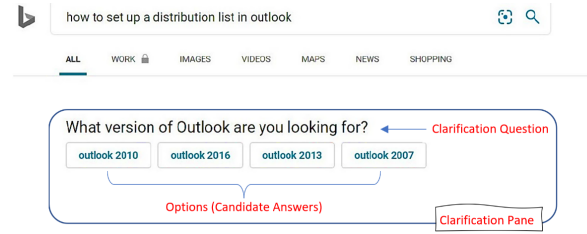

Figure 1: Example clarification pane from Bing [79], illustrating multi-choice options for an ambiguous query.

- **Task 1: List-wise Preference Rating** — Models simultaneously evaluate multiple clarification panes[2] and rank them to assess their *relative preferences*, evaluating the model's ability to handle comparative judgements and contextual reasoning.
- **Task 2: Pair-wise Quality Labelling** — Models evaluate the *overall quality* of a single query-clarification pair. Given individual annotation units, models assign holistic quality ratings, capturing their capacity for integrating multiple subjective criteria (e.g., clarity, relevance, usefulness) into a single judgement.
- **Task 3: Pair-wise Aspect Labelling** — Models evaluate multiple, fine-grained quality dimensions (e.g., *coverage*, *diversity*, *option order*) of a single query-clarification pair to yield interpretable, criteria-specific judgements.

## 3 Related Work

This review summarises previous research on the role of LLMs in annotation tasks, emphasising validation needs and hybrid human-LLM systems.

**Assessing LLM Annotation Performance.** LLMs have demonstrated mixed performance in annotation tasks [4, 23, 33, 46, 63]. They achieve human-level accuracy in relevance labelling [4], zero-shot and few-shot tasks via prompt engineering [82], cognitive event segmentation [12], and fluency in natural language generation evaluation [70]. HuggingChat and FLAN even outperform crowd workers in certain contexts [7]. However, these models remain vulnerable to adversarial inputs and systematic errors [4].

LLMs significantly reduce annotation costs, particularly in high-volume tasks [17, 22, 72], achieving comparable or superior accuracy to human annotators [22, 66]. However, they remain sensitive to prompt variations [26], struggle with complex language tasks [16, 33, 63], and show limitations in subjective and nuanced evaluations [40, 80, 83].

LLMs struggle with fine-grained distinctions [29, 47, 49, 81]. Studies show challenges in processing controversial sentence pairs [29], inconsistencies in human sciences [47], and difficulties with complex reasoning tasks in domains like oncology physics [32], sentiment analysis [12, 31], and legal annotation [57], where fine-tuned models often outperform GPT-4.

LLMs often exhibit bias and inconsistency [1, 10, 50, 52]. This includes relevance misalignment [1], demographic bias in refugee annotations [10], and positional bias in scoring [71]. Risks of misinformation and hallucinations further highlight their limited nuance [25], underscoring the need for improved alignment via Chain-of-Thought reasoning [44] and open-source initiatives [38].

---

[2]A clarification pane is defined as a multi-choice clarification question and its options.

**Applications of Hybrid Human-LLM Systems.** Due to challenges like bias and limited nuance, researchers have urged caution in fully automating LLM-based assessments [18, 45], advocating for HITL systems. Proposed HITL strategies include dynamic trust scores [58], adaptive thresholds [14], and diversity entropy thresholds [13]. Confidence-based methods have also been suggested, with thresholds of 95% [37], 70% [3], and dataset-specific adjustments [56].

Other studies support hybrid approaches. Li et al. [42] proposed using entropy-based methods to assign simpler tasks to LLMs and complex ones to humans. Wang et al. [74] explored reinforcement learning with human feedback. Human oversight was shown to improve label consistency [27, 64] and reduce errors in complex tasks [30, 51]. Soboroff [59] cautioned against using LLMs as ground truth generators, advocating for their use as collaborative tools to support human judgement and ethical oversight.

**Knowledge Gaps.** While LLMs perform well in general annotation tasks [27, 74], their effectiveness in high-cognitive-load tasks involving subjective and nuanced judgements is still underexplored [33, 80]. Our work addresses these limitations and offers a HITL-based approach that balances scalability and quality. It enhances alignment with human judgement while reducing annotation costs—crucial for high-stakes domains.

## 4 Methodology and Experimental Framework

### 4.1 Dataset

We use the *MIMICS-Duo* dataset [62], an extension of the *MIMICS* dataset [79], that contains real Bing queries, each with multiple clarification panes and rich human annotations across three dimensions: *preference*, *quality*, and *aspect-based evaluation*–performed by crowdsourcing workers on Amazon's Mechanical Turk (AMT). These diverse and subjective labels make them ideal for assessing LLMs on complex annotation tasks beyond binary relevance. Although created before the LLM era, *MIMICS-Duo* offers a unique advantage: its human annotations are entirely independent of LLM outputs, enabling a clean, unbiased comparison between LLM and human performance. While newer datasets exist, *MIMICS-Duo* remains one of the few publicly available resources combining ambiguous queries with multi-choice clarifications and fine-grained judgements. This makes it a strong fit for benchmarking our HITL pipeline and evaluating model generalisation under low-supervision settings. Table 1 shows dataset statistics.

**Table 1: Statistics of the *MIMICS-Duo* dataset [62].**

| | |
|---|---|
| Number of queries | 306 |
| Number of query-clarification pair | 1,034 |
| Number of clarifications per query | 3.38±0.68 |
| Min & max clarifications per query | 3 & 8 |
| Number of options | 3.59±1.2 |
| Min & max number of options | 2 & 5 |

### 4.2 Task Design

Following Tavakoli et al. [62], we structure our methodology into three independent labelling tasks. Each model receives the same annotation guidelines as those provided to crowdsourcing workers. Tasks are conducted separately to ensure isolation and avoid cross-task influence. Our aim is to assess how well LLMs replicate human judgements. We benchmark model outputs against the majority-voted labels from *MIMICS-Duo*, which serve as a practical proxy for human consensus. While human annotations are not perfect [69], majority voting helps mitigate individual bias. Our broader objective is to evaluate whether LLMs can reliably reduce or replace manual effort in a complex IR annotation pipeline. High agreement with human labels thus indicates potential for scalable automation.

*4.2.1* **Task 1: List-wise Preference Rating.** We provide the LLMs with a query along with all clarification panes generated for it. Then, the models are tasked to assign a rating on a 5-star scale (five means the highest preference, and one means the lowest preference) to each clarification pane. To simulate the human annotation process, all generated clarification panes for a given query are shown simultaneously, allowing the LLMs to rate them relatively. Like human annotators, LLMs can assign identical preference ratings to different clarification panes if appropriate.

*4.2.2* **Task 2: Pair-wise Quality Labelling.** The LLMs are provided with one query-clarification pair, and then they are tasked to rate the overall quality of the given clarification pane on a 5-level scale (1: very bad, 2: bad, 3: fair, 4: good, 5: very good).

*4.2.3* **Task 3: Pair-wise Aspect Labelling.** The LLMs are provided with a query-clarification pair, and then they are tasked with evaluating three aspects of the clarification pane: Coverage (i.e., the extent to which the clarification pane covers every potential aspect of the query), Diversity (i.e., the extent to which the clarification pane does not contain redundant information), and Options Order (i.e., the extent to which the most relevant and important options are positioned from left to right) on a scale from 1 to 5 based on the query.

### 4.3 Model Selection and Rationale

To evaluate how well LLMs perform annotation with minimal supervision, we select a diverse set of models, balancing performance, explainability, and transparency. We include GPT-4o for its strong reasoning and contextual understanding, and claude-3-haiku-20240307 for its high-quality, explainable outputs. We also use Cohere Command R+ (command-large-nightly), optimised for retrieval and reasoning tasks, aligning well with IR annotation needs—all three models representing powerful proprietary, black-box models.

To ensure reproducibility and transparency, we add Mistral 7B, a lightweight open-source transformer with dense attention and 7 billion parameters. Despite its smaller size, it performs comparably to larger models like Llama 2 [67], though it may underperform on tasks requiring deep reasoning [34, 48]. This range of models allows us to systematically assess annotation quality across proprietary and open models under minimal supervision settings.

*4.3.1* **Impact of Few-Shot Examples on LLM's Annotation Quality.** To investigate the impact of few-shot examples on LLM annotation performance, we employ a two-stage labelling process. In the first round, the models are provided only with task definitions, without any examples–a setup referred to as the *Zero-Shot* Setting (ZSS). In the second round, the models are presented with examples illustrating how crowdsourcing workers labelled different tasks–a setup referred to as the *Few-Shot* Setting (FSS). For instance, in Task 2, the models are shown five examples, representing quality labels from 1 to 5, assigned to five different query-clarification pairs.

*4.3.2 Temperature Settings and Prompt Sensitivity Analysis.* To investigate the sensitivity of LLMs to temperature settings across various labelling tasks, we calculate the standard deviation (SD) and entropy to measure uncertainty or randomness in the models and the consistency of outputs across different temperature settings (i.e., 0, 0.5, and 1). The temperature parameter in language models controls the uncertainty or creativity of the generated output. Lower values (e.g., 0.2) result in near-deterministic and focused responses. Higher values (e.g., 0.8) produce more varied and creative outputs. By adjusting the temperature setting, we aim to determine how response consistency and annotation quality might vary in relation to this parameter.

To investigate the impact of prompt sensitivity on different annotation tasks, we use LLMs to perform the annotations while varying the prompt configurations. Specifically, we rephrase the prompt by changing its structure or wording, alter the order of examples within the prompt, vary the prompt length by using shorter versions, and change the maximum token limit from the initial limit of 1000 to 250 and 2000. This systematic approach allows us to assess how these changes influence the models' performance in replicating annotation tasks.

## 4.4 Integrating A Human-in-the-Loop Workflow

Inspired by Rouzegar and Makrehchi [56], we employ confidence-based thresholding, with thresholds adjusted based on task complexity and resources. Higher thresholds are suited for high-stakes domains (e.g., medical or legal), while lower ones enable greater automation in less critical tasks. In our setting, the HITL workflow is valuable when models struggle with subtle distinctions (e.g., "good" vs. "very good"). **Our goal is to determine an optimal confidence threshold using a small annotated subset—minimising overall human effort while maintaining annotation quality.** The workflow proceeds as follows. Step 1 to Step 6 are followed on a subset of the dataset and Final Step is followed on the remaining dataset.

**Step 1. Manual annotation baseline.** The process starts with human annotation on a small dataset subset, with sample size determined by the dataset's scale. To illustrate how this workflow operates, we apply our HITL system to random samples of 5%, 10%, and 15% of the dataset, showing that 10% suffices.

**Step 2. LLM predictions on the human-labelled subset.** The same subset annotated by humans is also labelled by each LLM, which is run once per model. Alongside each prediction, the model outputs a confidence score. Confidence can be estimated in various ways, such as softmax probabilities, token-level scores, or verbalised self-assessments [60]. In our study, we adopt the verbalised confidence method, shown by Tian et al. [65] to be better calibrated than raw probabilities across models like ChatGPT, GPT-4, and Claude, and datasets such as TriviaQA [35], SciQ [75], and TruthfulQA [43].

**Step 3. Ensemble labelling.** Model-generated labels are aggregated using majority voting, mirroring the "collective wisdom" approach widely adopted in human annotation and used in Tavakoli et al. [62]. Rather than running a single model multiple times, we leverage multiple diverse LLMs, each contributing a label and an associated verbalised confidence score. These confidence scores are averaged, and their standard deviations are calculated to estimate a final mean confidence score and to quantify inter-model disagreement. This standard deviation acts as a proxy for uncertainty: high variance indicates low consensus among models and flags the instance for human review. This ensemble strategy allows us to examine whether aggregating predictions from diverse LLMs—similar to aggregating human annotators' votes—can better replicate human judgement than relying on a single model alone.

We recognise that confidence estimation complements uncertainty estimation by indicating how certain a model is in its own prediction. For models like GPT-4o and Claude 3, these scores may reflect internal probabilities, token alignment, rating scale adherence, or prompt ambiguity. As prior work shows [11, 53, 60, 76], these confidence mechanisms vary significantly across models due to differences in architecture, training data, and optimisation goals. Importantly, this variation is not a limitation—it reflects the diversity found in human annotation, where annotators bring different levels of expertise, experience, and interpretation. By aggregating across models, we aim to capture this diversity and use it to inform a robust, scalable human-in-the-loop annotation strategy.

**Step 4. Defining confidence and uncertainty thresholds.** Here, each LLM-aggregated label is paired with a mean confidence score and a corresponding standard deviation, reflecting both model agreement and uncertainty. Using the observed minimum and maximum values of these scores across each task, we define a range of *candidate confidence score thresholds* and *candidate standard deviation thresholds*. For example, if the observed mean confidence scores range from 70% to 100%, and standard deviations range from 0 to 12, we test confidence thresholds across 70–100% with a 5% interval and vary the standard deviation thresholds in increments of 2 within the observed bounds.[3] These thresholds allow us to simulate a variety of HITL configurations. Importantly, this thresholding strategy is entirely model- and data-driven—calibrated using model outputs rather than relying on human-labelled ground truth.

In the next steps, we apply each combination of candidate confidence score and standard deviation thresholds to identify which labels should be flagged for human review. We then evaluate the trade-off between annotation quality and human effort using a range of evaluation metrics and Pareto front optimisation [21] technique. This process allows us to identify the optimal threshold pair that minimises human involvement while maintaining high annotation quality. In other words, the HITL workflow is executed multiple times on the annotated subset, once for each threshold combination, to determine the most effective configuration.

**Step 5. Label acceptance via HITL rules.** A set of rules is implemented in the HITL system to determine whether an aggregated label is automatically accepted or flagged for human review. This process is conducted on every combination of *candidate confidence score* and *standard deviation thresholds*:

- **High Confidence and Low Variance:** If the mean confidence score ($\overline{c}_{LLMs}$) meets or exceeds the predefined confidence threshold ($c_{Threshold}$) and the standard deviation ($SD_{conf}$) is less than or equal to the threshold ($SD_{Threshold}$), the aggregated label ($\overline{y}_{LLMs(agg)}$) is automatically accepted.
- **High Confidence but High Variance:** If the mean confidence score meets or exceeds the threshold but the standard deviation exceeds the threshold, the label is flagged for human review. In this study, the aggregated label is replaced by the human-verified label obtained from the initial annotation on the subset.
- **Low Confidence:** If the mean confidence score falls below the threshold, regardless of the standard deviation, the label is flagged for human review.[4] The aggregated label is replaced by the human-verified label obtained from the initial annotation.

---

[3]A standard deviation of 2 indicates that, on average, the model-reported confidence scores for a given label deviate by two units from the mean confidence across models.

Figure 2 shows the modular pipeline implemented in this study.

**Step 6. Identifying optimal thresholds.** For each combination of confidence score and standard deviation thresholds, we compute key evaluation metrics—described in the following subsection—based on the updated aggregated labels. These metrics are assessed alongside the corresponding human effort, defined as the proportion of labels flagged for human review relative to the total number of labels. The optimal threshold pair is identified using Pareto front optimisation, aiming to maximise annotation quality while minimising human involvement. A constraint is applied to ensure this trade-off remains balanced, yielding the most practical configuration—referred to as the best solution.

While multiple metrics are calculated in this study, we use a Quadratically Weighted Cohen's Kappa (Kw) [19] as the primary evaluation metric, with a minimum acceptable threshold of 0.7, in line with commonly adopted standards for substantial inter-rater agreement in annotation studies [39]. This choice reflects our goal of achieving high agreement between the HITL-generated labels and human annotations. The selection of metric and its threshold is guided by human resource limitations and the required annotation quality for the task.

**Final Step. Applying the optimal HITL strategy to the dataset.** All preceding steps are conducted on the sampled subset to identify the optimal confidence score and standard deviation thresholds. Once these thresholds are determined, we apply the HITL workflow to the remainder of the dataset in a single pass using LLMs.

To simulate the HITL process, we leverage the existing human annotations from the *MIMICS-Duo* dataset, which were determined by majority voting among three crowdworkers. For any instance flagged for human review, we substitute the aggregated model label with the corresponding pre-existing human label—emulating a realistic workflow similar to using human annotators via platforms like AMT. In a practical deployment, such labels would instead be reviewed by live annotators. By repeating this process for each annotation task, we observe task-specific optimal thresholds, leading to varying levels of human effort reduction depending on the complexity and difficulty of the task for LLMs.

### 4.5 Evaluation Metrics
We use a comprehensive set of metrics to evaluate LLM performance across the three labelling tasks, focusing on accuracy, reliability, and robustness. The *MIMICS-Duo* dataset is inherently imbalanced—for example, in Task 1 (Preference Rating), labels 1 and 2 account for just 2.4% and 8.7% of instances, while labels 3–5 dominate. To avoid misleading results from standard accuracy metrics, we calculate class-wise Precision and F1-Score to assess model behaviour on minority labels. We then compute and report Macro Precision and Macro F1-Score, which are more suitable for imbalanced data.

To assess agreement with human annotations while accounting for disagreement severity, we use Quadratically Weighted Cohen's Kappa (Kw), which penalises larger label mismatches more heavily. For error analysis, we report Mean Absolute Error (MAE), which measures the average absolute difference between predicted and true labels, offering a robust, outlier-resistant view of model error. We also include Pearson's correlation coefficient ($\rho$) to capture linear trends in agreement, complementing other metrics despite the ordinal nature of the labels.

To analyse error patterns, we use confusion matrices to visualise class-by-class prediction distributions and identify systematic misclassifications (e.g., frequent confusion between labels 3 and 4). We also report Confidence-Weighted Accuracy (CWA) [15], which combines prediction correctness with confidence calibration—rewarding high-confidence correct predictions and penalising overconfident errors. It is defined as:

$$CWA = \frac{\sum(\text{Confidence of Correct Predictions})}{\sum(\text{Confidence of all Predictions})} \quad (1)$$

While confidence scores are normalised to 0–100%, their interpretation varies across models (e.g., 90% from GPT-4o may not match 90% from Claude 3) due to differences in architecture and confidence estimation. CWA accounts for these disparities by weighting prediction correctness by each model's reported confidence.

To estimate how much annotation effort our HITL framework can save, we define and compute the Human Effort Reduction (*HER*) factor. This metric captures the percentage of instances where LLM predictions are confident enough to bypass human review, based on defined thresholds. It is defined as:

$$HER = (1 - \frac{\text{No. of Instances Sent for Human Review}}{\text{Total No. of Instances}}) * 100 \quad (2)$$

A higher *HER* reflects greater efficiency—fewer labels require manual review without sacrificing quality.

Beyond performance, we assess model robustness through temperature and prompt sensitivity analyses, using two uncertainty-aware metrics:

- **Entropy:** Measures uncertainty over predicted label distributions—higher values indicate greater indecision.
- **Standard Deviation:** Captures variability in predictions across different prompts or temperature settings, reflecting sensitivity to input changes.

These metrics reveal how stable model outputs are under varying conditions. Combined with traditional evaluation scores, they offer a more complete picture of LLM reliability—especially important when human oversight is limited.

## 5 Results and Analyses
### 5.1 Task 1: List-wise Preference Rating
We present model performance on Task 1 (Subsection 4.2.1) using a temperature setting of 0, which showed the highest alignment with crowdsourced labels. Although models were also tested at temperatures 0.5 and 1, this setting yielded the best results—consistent with prior findings [16, 27, 40, 66].

---
[4]In cases where the mean confidence score is low but the standard deviation is also low, the system interprets this as consistent but low-confidence predictions across models. Such cases are flagged for human review.

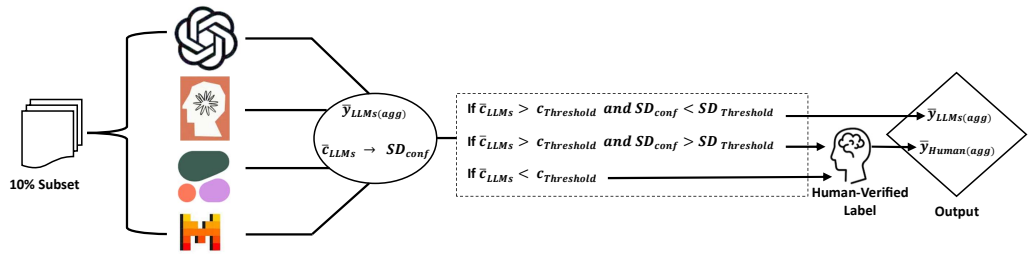

**Figure 2: HITL workflow, presented for one combination of the confidence score and standard deviation thresholds.**

As shown in Table 2, the FSS approach significantly improved Claude 3 and Cohere Command R, while having minimal impact on GPT-4o. In contrast, Mistral 7B's performance declined across most metrics except Macro Precision. This may stem from its smaller context window, which can limit its ability to process the extended prompt used in this task. Additionally, smaller models like Mistral 7B are more prone to overfitting with few-shot examples, reducing generalisation. Overall, GPT-4o, Claude 3, and Cohere performed comparably, with GPT-4o and Claude 3 slightly ahead—while Mistral 7B consistently underperformed.

To implement the HITL system (Subsection 4.4), we annotated a single random subset of the dataset. While only one sample is required to estimate optimal thresholds, we tested subsets of 5%, 10%, and 15% to examine the impact of sample size. Using individual LLMs, we generated aggregated labels with mean confidence scores and standard deviations. Figure 4 shows consistent trends across metrics, with greater variability at 5%. The close alignment between the 10% and 15% subsets suggests that 10% is sufficient for reliable threshold estimation—reducing human effort across tasks.

On the 10% subset in Task 1, confidence scores ranged from 66.3% to 98.8%, with standard deviations between 2.5 and 42.1. The optimal confidence score (90%) and standard deviation (14) thresholds were identified using Pareto front optimisation between weighted Cohen's Kappa (Kw) and human effort, as shown in Figure 3. Applying these thresholds in the HITL workflow significantly improved the automatic labelling process for Task 1, **reducing human effort by 45% while maintaining performance comparable to full human annotation** (see Table 2). It is worth noting that in our experiments, individual models can also serve as baselines, helping to demonstrate the effectiveness of HITL in achieving higher-quality annotations with reduced manual effort.

The confusion matrices of the models, as heatmaps, are shown in (Figure 5). The HITL model stood out with higher diagonal values, indicating superior accuracy and lower misclassification rates. While GPT-4o and Cohere Command R performed moderately, they showed noticeable overlaps between classes. Claude 3 and Mistral struggled more, with Mistral displaying the highest confusion and least reliable predictions. A common challenge across models was distinguishing adjacent classes (e.g., class 3 vs. 4, class 4 vs. 5), but the HITL system consistently minimised misclassification.

### 5.2 Task 2: Pair-wise Quality Labelling

We evaluated model performance on Task 2 (Section 4.2.2). As shown in Table 2, all models benefited from the few-shot setting,

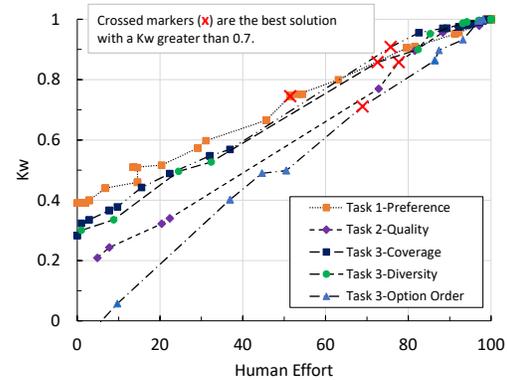

**Figure 3: Pareto Front for Kw vs. human effort for all Tasks on 10% Subset. Human effort is defined as the proportion of labels flagged for human review relative to the total number of labels.**

with Claude 3 seeing the largest gains, followed by GPT-4o. Claude 3 led in Macro Precision, Macro F1, and Kw, while GPT-4o ranked second overall. Cohere Command R showed weaker performance across most metrics. In the HITL workflow, mean confidence scores ranged from 75% to 96.3%, with standard deviations from 4.8 to 15. To find optimal thresholds, we tested confidence scores from 80%–96.3% (in 5% steps) and standard deviations in increments of 2 across three subsets (see Figure 3). The best-performing combination—80% confidence and a standard deviation of 10—was applied to the full dataset, improving all metrics and **reducing human effort by 26%** (see Table 2).

Figure 5 presents the confusion matrices for Task 2. The HITL system showed the fewest misclassifications, outperforming all individual models. GPT-4o captured some Class 4 instances but struggled with precision due to frequent errors. Claude 3 performed well on Class 4 but misclassified Classes 2, 3, and 5. Cohere Command R was heavily biased toward Classes 3 and 4, rarely predicting Classes 1, 2, or 5. Mistral showed similar bias, favouring Classes 1 and 3 while failing to predict Classes 4 and 5 accurately.

### 5.3 Task 3: Pair-wise Aspect Labelling

Here, the performance of the models in conducting Task 3, described in 4.2.3, is presented.

**Coverage Labelling.** Table 2 shows that GPT-4o improved across key metrics under FSS—Macro F1, MAE, Kw, Pearson correlation, and CWA—while Macro Precision remained stable, indicating it benefited from added context in Coverage Labelling. Claude 3 saw minimal change, with slight drops in consistency. Cohere Command R was negatively affected, with declines in most metrics and only minor gains in Macro Precision and MAE. Mistral 7B showed mixed results: improved Macro Precision but higher MAE and lower CWA, suggesting better class-specific precision but reduced overall reliability. In the HITL workflow, mean confidence scores ranged from 72.5% to 96.3%, with standard deviations from 0 to 22.9. We tested confidence thresholds from 75%–96.3% and standard deviation thresholds in steps of 2 across three subsets. The optimal configuration—75% confidence and a standard deviation of 12—was applied to the full dataset, improving all metrics and **reducing human effort by 26%** (see Table 2).

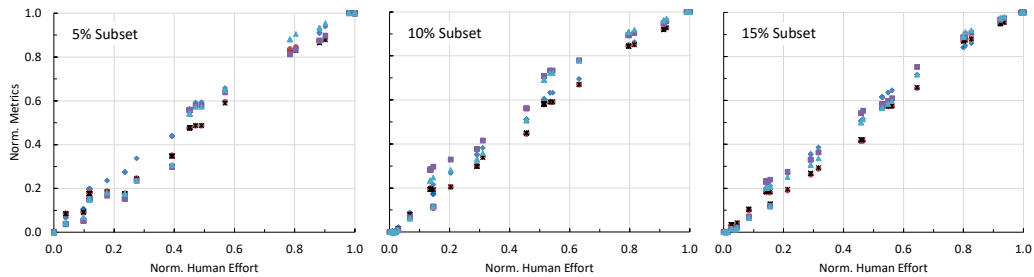

Figure 4: Normalised metrics vs. human effort when implementing HITL in annotating Task 1 on 5%, 10% and 15% subsets. Each marker represents results of a calculated metric (e.g., Macro Precision).

Table 2: Comparing performance of the models with the crowdsourcing workers across all annotation tasks under ZSS and FSS. The temperature setting for all models is 0. Preference refers to Task 1, and Quality refers to Task 2.

| | Annotator | Metric/Learning Approach | | | | | | | | | | | |
|---|---|---|---|---|---|---|---|---|---|---|---|---|---|
| | | **Macro Precision** | | **Macro F1-Score** | | **MAE** | | **Kw** | | **Pearson ($\rho$)** | | **CWA** | |
| | | ZSS | FSS | ZSS | FSS | ZSS | FSS | ZSS | FSS | ZSS | FSS | ZSS | FSS |
| Preference | GPT-4o | 0.343 | 0.338 | 0.294 | 0.301 | 0.836 | 0.865 | 0.313 | 0.330 | 0.324 | 0.362 | 0.361 | 0.355 |
| | Claude 3 | 0.281 | 0.497 | 0.253 | 0.290 | 0.882 | 0.793 | 0.197 | 0.315 | 0.201 | 0.366 | 0.345 | 0.398 |
| | Cohere Com. R | 0.161 | 0.344 | 0.148 | 0.314 | 1.191 | 0.907 | 0.017 | 0.246 | 0.018 | 0.258 | 0.264 | 0.402 |
| | Mistral 7B | 0.211 | 0.335 | 0.178 | 0.160 | 1.277 | 1.766 | 0.030 | 0.021 | 0.040 | 0.029 | 0.243 | 0.177 |
| | HITL with **45% HER** | | **0.765** | | **0.758** | | **0.307** | | **0.757** | | **0.758** | | **0.777** |
| Quality | GPT-4o | 0.230 | 0.260 | 0.183 | 0.230 | 1.131 | 1.015 | 0.182 | 0.224 | 0.286 | 0.306 | 0.283 | 0.333 |
| | Claude 3 | 0.259 | 0.342 | 0.199 | 0.307 | 1.032 | 0.665 | 0.183 | 0.255 | 0.271 | 0.274 | 0.356 | 0.477 |
| | Cohere Com. R | 0.198 | 0.179 | 0.173 | 0.203 | 0.531 | 0.543 | 0.088 | 0.143 | 0.120 | 0.176 | 0.519 | 0.527 |
| | Mistral 7B | 0.113 | 0.266 | 0.067 | 0.072 | 1.032 | 1.517 | 0.0 | 0.029 | 0.0 | 0.075 | 0.191 | 0.123 |
| | HITL with **26% HER** | | **0.729** | | **0.745** | | **0.174** | | **0.779** | | **0.787** | | **0.873** |
| Coverage | GPT-4o | 0.338 | 0.337 | 0.217 | 0.306 | 1.079 | 0.950 | 0.284 | 0.380 | 0.380 | 0.437 | 0.306 | 0.365 |
| | Claude 3 | 0.344 | 0.305 | 0.237 | 0.248 | 0.985 | 0.857 | 0.293 | 0.260 | 0.345 | 0.270 | 0.312 | 0.382 |
| | Cohere Com. R | 0.195 | 0.199 | 0.201 | 0.155 | 0.847 | 0.784 | 0.199 | 0.112 | 0.228 | 0.178 | 0.366 | 0.401 |
| | Mistral 7B | 0.284 | 0.611 | 0.156 | 0.161 | 0.790 | 1.092 | 0.041 | 0.094 | 0.077 | 0.192 | 0.421 | 0.202 |
| | HITL with **26% HER** | | **0.833** | | **0.794** | | **0.239** | | **0.844** | | **0.849** | | **0.840** |
| Diversity | GPT-4o | 0.219 | 0.256 | 0.231 | 0.258 | 1.149 | 1.018 | 0.263 | 0.313 | 0.277 | 0.319 | 0.326 | 0.354 |
| | Claude 3 | 0.246 | 0.287 | 0.232 | 0.288 | 1.026 | 0.955 | 0.199 | 0.250 | 0.217 | 0.252 | 0.345 | 0.368 |
| | Cohere Com. R | 0.199 | 0.208 | 0.153 | 0.138 | 0.900 | 0.878 | 0.127 | 0.111 | 0.159 | 0.148 | 0.299 | 0.302 |
| | Mistral 7B | 0.191 | 0.294 | 0.138 | 0.092 | 0.811 | 1.111 | 0.055 | 0.017 | 0.103 | 0.061 | 0.383 | 0.167 |
| | HITL with **24% HER** | | **0.851** | | **0.846** | | **0.221** | | **0.847** | | **0.849** | | **0.855** |
| Option Order | GPT-4o | 0.339 | 0.300 | 0.254 | 0.267 | 0.972 | 0.990 | 0.163 | 0.129 | 0.201 | 0.147 | 0.349 | 0.360 |
| | Claude 3 | 0.211 | 0.231 | 0.173 | 0.163 | 1.039 | 0.876 | 0.047 | 0.049 | 0.047 | 0.059 | 0.277 | 0.322 |
| | Cohere Com. R | 0.162 | 0.241 | 0.118 | 0.151 | 1.168 | 1.125 | 0.039 | 0.057 | 0.082 | 0.096 | 0.203 | 0.229 |
| | Mistral 7B | 0.210 | 0.221 | 0.199 | 0.137 | 1.003 | 0.753 | 0.077 | 0.014 | 0.081 | 0.054 | 0.277 | 0.403 |
| | HITL with **30% HER** | | **0.892** | | **0.836** | | **0.226** | | **0.794** | | **0.799** | | **0.840** |

**Diversity Labelling.** GPT-4o showed consistent improvement across all metrics under FSS, while Claude 3 performed similarly with notable gains. In contrast, Cohere Command R and Mistral 7B struggled, particularly with capturing diversity. Mistral 7B showed the weakest alignment with human annotations, with the lowest Kw and Pearson correlation. Its CWA dropped under FSS likely reflects overconfident errors, poor generalisation, and limited capacity. In the HITL workflow, mean confidence scores ranged from 77.5% to 100%, with standard deviations between 0 and 16.5. Testing thresholds from 80% to 100%, we identified 80% confidence and a standard deviation of 12 as optimal (Figure 3). Applying these thresholds to the full dataset yielded comparable quality while **reducing human effort by 24%** (see Table 2).

**Option Order Labelling.** GPT-4o consistently outperformed other models across most metrics, though FSS offered little benefit and

**Task 1: List-wise Preference Rating**

| | P1 | P2 | P3 | P4 | P5 | P1 | P2 | P3 | P4 | P5 | P1 | P2 | P3 | P4 | P5 | P1 | P2 | P3 | P4 | P5 | P1 | P2 | P3 | P4 | P5 |
|---|---|---|---|---|---|---|---|---|---|---|---|---|---|---|---|---|---|---|---|---|---|---|---|---|---|
| T1 | 6 | 11 | 4 | 4 | 0 | 2 | 6 | 8 | 6 | 3 | 10 | 4 | 4 | 6 | 1 | 9 | 10 | 0 | 1 | 5 | 18 | 3 | 1 | 3 | 0 |
| T2 | 6 | 35 | 31 | 17 | 1 | 0 | 8 | 25 | 48 | 9 | 13 | 14 | 25 | 32 | 6 | 21 | 37 | 3 | 3 | 26 | 2 | 71 | 5 | 11 | 1 |
| T3 | 9 | 74 | 126 | 91 | 13 | 0 | 11 | 59 | 170 | 73 | 29 | 26 | 128 | 113 | 17 | 89 | 98 | 24 | 6 | 96 | 1 | 12 | 221 | 65 | 14 |
| T4 | 1 | 47 | 114 | 166 | 19 | 0 | 4 | 24 | 208 | 111 | 24 | 16 | 97 | 189 | 21 | 111 | 91 | 6 | 24 | 115 | 3 | 4 | 15 | 307 | 18 |
| T5 | 1 | 23 | 72 | 133 | 30 | 0 | 1 | 17 | 113 | 128 | 17 | 6 | 71 | 103 | 62 | 78 | 83 | 5 | 4 | 89 | 2 | 4 | 11 | 64 | 178 |
| | | *GPT-4o* | | | | | *Claude 3* | | | | | *Cohere Command R* | | | | | *Mistral 7B* | | | | | *HITL* | | | |

**Task 2: Pair-wise Quality Labelling**

| | P1 | P2 | P3 | P4 | P5 | P1 | P2 | P3 | P4 | P5 | P1 | P2 | P3 | P4 | P5 | P1 | P2 | P3 | P4 | P5 | P1 | P2 | P3 | P4 | P5 |
|---|---|---|---|---|---|---|---|---|---|---|---|---|---|---|---|---|---|---|---|---|---|---|---|---|---|
| T1 | 2 | 1 | 1 | 0 | 0 | 1 | 0 | 1 | 2 | 0 | 0 | 0 | 1 | 3 | 0 | 3 | 0 | 1 | 0 | 0 | 4 | 0 | 0 | 0 | 0 |
| T2 | 6 | 16 | 6 | 5 | 0 | 2 | 11 | 7 | 12 | 1 | 0 | 0 | 14 | 19 | 0 | 9 | 1 | 23 | 0 | 0 | 3 | 27 | 1 | 2 | 0 |
| T3 | 21 | 86 | 41 | 43 | 10 | 0 | 45 | 56 | 93 | 7 | 0 | 1 | 71 | 129 | 0 | 65 | 3 | 133 | 0 | 0 | 4 | 5 | 167 | 25 | 0 |
| T4 | 12 | 140 | 113 | 221 | 78 | 0 | 56 | 79 | 390 | 39 | 0 | 1 | 98 | 465 | 0 | 131 | 2 | 428 | 3 | 0 | 3 | 2 | 31 | 522 | 6 |
| T5 | 8 | 47 | 30 | 93 | 54 | 0 | 20 | 16 | 170 | 26 | 0 | 0 | 36 | 196 | 0 | 56 | 1 | 175 | 0 | 0 | 3 | 1 | 12 | 45 | 171 |
| | | *GPT-4o* | | | | | *Claude 3* | | | | | *Cohere Command R* | | | | | *Mistral B7* | | | | | *HITL* | | | |

**Task 3: Coverage Labelling**

| | P1 | P2 | P3 | P4 | P5 | P1 | P2 | P3 | P4 | P5 | P1 | P2 | P3 | P4 | P5 | P1 | P2 | P3 | P4 | P5 | P1 | P2 | P3 | P4 | P5 |
|---|---|---|---|---|---|---|---|---|---|---|---|---|---|---|---|---|---|---|---|---|---|---|---|---|---|
| T1 | 7 | 20 | 3 | 1 | 0 | 1 | 8 | 8 | 13 | 1 | 0 | 0 | 9 | 22 | 0 | 2 | 0 | 28 | 1 | 0 | 23 | 0 | 7 | 1 | 0 |
| T2 | 15 | 81 | 29 | 20 | 0 | 1 | 27 | 44 | 70 | 3 | 0 | 0 | 44 | 101 | 0 | 1 | 13 | 128 | 3 | 0 | 1 | 104 | 36 | 4 | 0 |
| T3 | 5 | 44 | 36 | 37 | 4 | 0 | 14 | 22 | 81 | 9 | 0 | 0 | 27 | 98 | 1 | 0 | 0 | 124 | 2 | 0 | 0 | 5 | 106 | 14 | 1 |
| T4 | 17 | 112 | 96 | 184 | 38 | 3 | 46 | 82 | 269 | 47 | 0 | 0 | 67 | 378 | 2 | 1 | 0 | 395 | 45 | 6 | 0 | 10 | 58 | 377 | 2 |
| T5 | 2 | 48 | 37 | 135 | 63 | 0 | 19 | 25 | 175 | 66 | 0 | 0 | 29 | 254 | 2 | 1 | 0 | 259 | 7 | 18 | 0 | 0 | 31 | 22 | 232 |
| | | *GPT-4o* | | | | | *Claude 3* | | | | | *Cohere Command R* | | | | | *Mistral 7B* | | | | | *HITL* | | | |

**Task 3: Diversity Labelling**

| | P1 | P2 | P3 | P4 | P5 | P1 | P2 | P3 | P4 | P5 | P1 | P2 | P3 | P4 | P5 | P1 | P2 | P3 | P4 | P5 | P1 | P2 | P3 | P4 | P5 |
|---|---|---|---|---|---|---|---|---|---|---|---|---|---|---|---|---|---|---|---|---|---|---|---|---|---|
| T1 | 4 | 6 | 0 | 3 | 2 | 4 | 6 | 0 | 3 | 2 | 0 | 0 | 9 | 6 | 0 | 1 | 0 | 14 | 0 | 0 | 14 | 0 | 1 | 0 | 0 |
| T2 | 19 | 60 | 9 | 60 | 25 | 9 | 37 | 22 | 84 | 21 | 0 | 0 | 91 | 82 | 0 | 0 | 0 | 170 | 3 | 0 | 0 | 143 | 11 | 15 | 4 |
| T3 | 9 | 39 | 3 | 55 | 50 | 1 | 18 | 18 | 76 | 43 | 0 | 0 | 55 | 101 | 0 | 0 | 0 | 151 | 4 | 1 | 0 | 4 | 130 | 14 | 8 |
| T4 | 22 | 83 | 9 | 143 | 158 | 9 | 60 | 30 | 196 | 120 | 0 | 0 | 160 | 254 | 1 | 1 | 0 | 397 | 16 | 1 | 1 | 3 | 29 | 350 | 32 |
| T5 | 6 | 27 | 4 | 83 | 155 | 3 | 24 | 13 | 113 | 122 | 0 | 0 | 81 | 193 | 1 | 0 | 0 | 264 | 10 | 1 | 0 | 2 | 19 | 26 | 228 |
| | | *GPT-4o* | | | | | *Claude 3* | | | | | *Cohere Command R* | | | | | *Mistral B7* | | | | | *HITL* | | | |

**Task 3: Option Order**

| | P1 | P2 | P3 | P4 | P5 | P1 | P2 | P3 | P4 | P5 | P1 | P2 | P3 | P4 | P5 | P1 | P2 | P3 | P4 | P5 | P1 | P2 | P3 | P4 | P5 |
|---|---|---|---|---|---|---|---|---|---|---|---|---|---|---|---|---|---|---|---|---|---|---|---|---|---|
| T1 | 3 | 2 | 9 | 1 | 1 | 0 | 0 | 6 | 10 | 0 | 0 | 0 | 1 | 8 | 7 | 1 | 0 | 15 | 0 | 0 | 15 | 0 | 0 | 1 | 0 |
| T2 | 23 | 32 | 56 | 13 | 9 | 0 | 9 | 16 | 104 | 4 | 0 | 1 | 15 | 76 | 41 | 1 | 1 | 131 | 0 | 0 | 0 | 89 | 16 | 27 | 1 |
| T3 | 42 | 58 | 248 | 36 | 32 | 8 | 17 | 44 | 334 | 13 | 0 | 2 | 50 | 191 | 173 | 1 | 1 | 414 | 0 | 0 | 0 | 3 | 363 | 47 | 3 |
| T4 | 25 | 66 | 151 | 68 | 17 | 2 | 18 | 26 | 273 | 8 | 0 | 0 | 39 | 129 | 159 | 1 | 0 | 326 | 0 | 0 | 0 | 1 | 34 | 289 | 3 |
| T5 | 14 | 21 | 52 | 25 | 30 | 1 | 5 | 11 | 118 | 7 | 0 | 0 | 7 | 68 | 67 | 1 | 0 | 140 | 1 | 0 | 0 | 0 | 14 | 35 | 93 |
| | | *GPT-4o* | | | | | *Claude 3* | | | | | *Cohere Command R* | | | | | *Mistral B7* | | | | | *HITL* | | | |

**Figure 5: Comparison LLMs and HITL system predictions (P1,...,P5) against the crowdsourcing workers' labels (T1,...,T5) across all annotation Tasks. Results are based on the temperature setting value of zero.**

even hurt performance. Similar patterns were observed for Claude 3 and Mistral 7B, while Cohere Command R was the only model to benefit from FSS in this task. This may be because models like GPT-4o and Claude 3, with extensive pre-training, already performed well on ordering tasks—making few-shot examples unnecessary or prone to overfitting. Additionally, the subtlety of option order evaluation may be hard to capture with limited examples. In the HITL workflow, mean confidence scores ranged from 70% to 96.3%, with standard deviations between 0 and 15. The optimal thresholds—75% confidence and a standard deviation of 6—identified in Figure 3, were applied to the full dataset, achieving comparable annotation quality while **reducing human effort by 30%** (see Table 2).

Confusion matrices (Figure 5) show that the HITL system achieved the best precision and recall, with minimal misclassification. GPT-4o and Claude 3 performed reasonably well but struggled with adjacent class boundaries. Cohere Command R and Mistral underperformed, with Mistral showing the weakest results due to widespread confusion. Cohere is biased toward Labels 4 and 5, while Mistral heavily favoured Class 3 and generalises poorly across other classes.

To calibrate the confidence and agreement thresholds for the HITL pipeline, we reserved 10% of the dataset as a pre-annotation set for Pareto analysis, which was excluded from the final evaluation. For the remaining 90%, we computed the *HER*, which ranged from 24% to 45% depending on the task—indicating that up to half of the data can be automatically labelled with minimal quality loss. Even after factoring in the 10% setup cost, the overall reduction in manual effort remained substantial.

### 5.4 When Human Oversight is Most Needed

To better understand the conditions under which human oversight is most critical, we conducted a domain-level analysis of queries flagged for review by our HITL workflow. Our goal was to identify whether certain types of queries or topics were more prone to low-confidence or inconsistent LLM predictions. Inspired by Li et al. [41], we used GPT-4o alongside regular expressions and lexical cues to classify queries into domain categories, including *Technical Topics*, *Entertainment*, *Vehicles*, *Property*, *Nutrition/Health/Fitness*, *Religious References*, *Legal/Tax*, *Shopping/E-commerce*, and *Travel/Maps*. This classification was based on the semantic context of each query and its associated clarification options. The domain categories were selected to maximise the number of queries represented in each, enabling more robust and meaningful analysis. Queries that did not match any predefined patterns were noted as *Other*.

A key observation across all tasks was the consistently large size of the "Other" category, even after multiple refinements, indicating that a significant proportion of queries were highly diverse, context-dependent, or semantically ambiguous. Structured domains like Technical Topics and Entertainment frequently emerged among the most challenging for models, especially when combined with vague query intent or poorly differentiated options. Tasks like Coverage and Diversity triggered more domain ambiguity, while Preference and Option Order posed difficulties in subjective ranking, suggesting that the complexity of the annotation task directly influenced model performance. These patterns highlighted limitations in LLMs' ability to generalise across multi-dimensional judgement tasks.

Our analysis revealed that LLMs struggle particularly in domains requiring either contextual disambiguation or subjective reasoning. For example, in the Option Order task, the query "iphone screen repair" with options like "Apple store", "Nearby service centre", and "Mail-in repair" presented subtle ranking challenges. Models often disagreed on which option should appear first, especially when confidence was high but variance across models was significant. Similarly, in the Coverage task, queries like "types of yoga poses" led to inconsistent labelling when options overlapped in content scope but varied in phrasing, such as "beginner-friendly yoga" versus "restorative poses". These inconsistencies reinforced the value of the HITL workflow, where flagged cases were effectively isolated for targeted human intervention, ensuring high-quality labels without requiring full human annotation of the dataset.

Table 3: Temperature and prompt sensitivity analysis.

|  | GPT-4o | | Claude 3 | | Cohere Com. R | | Mistral 7B | |
| --- | --- | --- | --- | --- | --- | --- | --- | --- |
| Task | Entr.§ | SD§ | Entr. | SD | Entr. | SD | Entr. | SD |
| *Preference†* | 0.582 | 0.337 | 0.335 | 0.176 | 0.539 | 0.387 | 0.658 | 0.718 |
| *Quality†* | 0.360 | 0.187 | 0.255 | 0.131 | 0.124 | 0.064 | 0.991 | 0.775 |
| *Coverage* | 0.357 | 0.185 | 0.251 | 0.129 | 0.208 | 0.107 | 0.904 | 0.602 |
| *Diversity* | 0.438 | 0.269 | 0.343 | 0.186 | 0.425 | 0.218 | 0.999 | 0.694 |
| *Option Order* | 0.475 | 0.293 | 0.245 | 0.132 | 0.459 | 0.270 | 1.072 | 0.761 |
| *Coverage‡* | 0.442 | 0.237 | 0.649 | 0.360 | 0.470 | 0.258 | 0.928 | 0.813 |

† Preference refers to Task 1 and Quality refers to Task 2.
§ Entr. stands for entropy and SD stands for standard deviation.
‡ Prompt sensitivity in conducting Coverage labelling using Temp. 0 and max token of 1000.

## 5.5 Sensitivity Analysis

Table 3 shows LLM sensitivity to temperature changes (i.e., 0, 0.5, 1) using entropy and standard deviation on the full *MIMICS-Duo* dataset under the Few-Shot Setting (FSS), which yielded the best overall performance. Claude 3 was the most stable, with the lowest entropy and variance. GPT-4o balanced consistency and adaptability, while Cohere Command R was more sensitive, and Mistral 7B was the most unstable. Task-wise, Preference and Option Order were most affected by temperature, while Diversity showed moderate sensitivity. These results highlight the importance of selecting models and tuning temperatures based on task needs—Claude 3 suits stable tasks, whereas GPT-4o offers a flexible, balanced choice.

Table 3 also indicates prompt sensitivity conducted for the Coverage labelling task. GPT-4o was the most robust, showing minimal variation, and was best suited for producing stable, consistent annotations. Claude 3 and Cohere Command R stroke a balance between consistency and flexibility, while Mistral 7B was the most sensitive—more suited to tasks tolerant of variability but less reliable for replicating human annotations. These results underscore the importance of model selection based on desired stability, adaptability, and alignment with human judgement. Due to space constraints, prompt sensitivity results for other tasks are available at https://github.com/Leila-Ta/HITL-Framework-Search-Clarification.

## 6 Discussion

Our experiments revealed that all LLMs struggled with the annotation tasks, regardless of complexity. However, GPT-4o and Claude 3 showed closer alignment with human annotations, likely due to broader pre-training and stronger contextual understanding. Still, both models faced challenges in nuanced tasks like Preference and Option Order, which required fine-grained distinctions. Mistral 7B and Cohere Command R performed less reliably, showing lower stability and higher entropy—likely due to limited pre-training and weaker capacity for modelling complex class relationships. Sensitivity to few-shot settings was also task-dependent. While GPT-4o and Claude 3 occasionally showed diminished or even negative returns on simpler tasks—possibly due to redundancy or overfitting, Cohere Command R sometimes benefited, likely needing more explicit guidance. Mistral 7B's high misclassification rates further highlight the limitations of smaller models in this task.

In contrast, our HITL framework—requiring tuning on just 10% of a human-annotated subset to identify optimal thresholds for full-scale deployment—outperforms all individual models. By integrating model predictions, confidence scores, and selective human input, it achieved stronger alignment with human consensus while significantly reducing manual effort. The HITL workflow addresses model disparities by aggregating their strengths and mitigating weaknesses. Through confidence thresholds, standard deviation filtering, and targeted human validation, it enhances annotation quality, reduced noise, and improved reliability across tasks. Human effort was reduced by 24–45% across tasks, demonstrating the framework's efficiency. Importantly, HITL is task-dependent. For example, in Coverage labelling, it achieved a Kw of 0.844 with a 26% reduction in effort. If a lower Kw of 0.718 is acceptable, effort reduction increases to 60%. This flexibility highlights the importance of defining task-specific quality thresholds.

Our findings also highlight the importance of using multiple evaluation metrics. While metrics like weighted Cohen's Kappa, Macro F1-Score, and Pearson correlation offer useful insights, they often fail to capture issues related to misclassifications, particularly in unbalanced datasets in the context of annotation tasks. For example, a model might show strong overall performance but disproportionately misclassify certain classes, as seen with Cohere Command R's bias toward Labels 3 and 4. These findings underscore the need for a holistic evaluation framework that combines multiple metrics to assess model performance more comprehensively.

Note that while our HITL framework does not involve real-time human interaction, it simulated human-in-the-loop decision-making using majority-voted crowdsourced labels from the *MIMICS-Duo* dataset as proxies for human review. Rather than replicating synchronous collaboration, our goal is to approximate when and where human input is needed.

## 7 Conclusions and Future Work

This study evaluated the ability of LLMs to replicate human annotations in search clarification tasks and introduced a HITL framework to improve annotation quality while reducing human effort. We observed that assessed LLMs in this study struggled with nuanced distinctions. The HITL approach addressed these gaps by aggregating model predictions, applying confidence-based filtering, and focusing human input on low-confidence cases—achieving comparable quality with up to 45% less human effort.

A limitation of this work is that it has been tested on a single dataset. Although this is considered acceptable based on ICTIR CFP, future work should study the generalisability of these findings across tasks and datasets. Future work could also extend this work by exploring adaptive learning and evaluating alternative ensemble strategies in HITL. Fine-tuning LLMs for specific tasks may also further improve alignment with human judgement and reduce reliance on manual review.

**Acknowledgements.** This work was supported in part by the Center for Intelligent Information Retrieval, in part by the Office of Naval Research contract number N000142212688, and in part by NSF grant number 2143434. Any opinions, findings and conclusions or recommendations expressed in this material are those of the authors and do not necessarily reflect those of the sponsor.